# A 0.5-V Linear Neuromorphic Voltage-to-Spike Encoder Using a Bulk-Driven Transconductor

Meysam Akbari
ZIAM & CogniGron
University of Groningen
9747 AG Groningen, The Netherlands
m.akbari@rug.nl

Erika Covi
ZIAM & CogniGron
University of Groningen
9747 AG Groningen, The Netherlands
e.covi@rug.nl

Kea-Tiong Tang
Department of Electrical Engineering
National Tsing Hua University
Hsinchu 30013, Taiwan
kttang@mx.nthu.edu.tw

*Abstract*—This work introduces an ultralow-power voltage-to-spike encoder that achieves near-linear voltage-to-firing-rate conversion by pairing a linearized bulk-driven transconductor with a DPI-based LIF neuron. A tail-less bulk-driven differential pair improves large-signal linearity, while a translinear linearization network suppresses the dominant *sinh* nonlinearity and stabilizes the bias-tunable V→I gain. The resulting current feeds a DPI front-end that linearizes current-to-spike conversion. Fabricated in TSMC 0.18-μm CMOS and operating at $V_{DD}$=0.5 V with 2–27 nA reference current, the encoder achieves a deviation of less than 5.6% from linearity over 0.1–0.4 V input, consumes 22–180 nW, and occupies 0.0074 mm².

*Keywords—Transconductor, Neuron, Encoder, Signal-to-spike, Linear, Neuromorphic, Firing, Biological.*

## I. Introduction

Neuromorphic sensors can convert analog signals directly into spikes, avoiding analog-to-digital converters (ADC) and reducing energy and area [1–2]. Designs using leaky integrate-and-fire (LIF) neurons or delta modulators encode the input amplitude in the firing rate, enabling operations like on-chip spectral analysis [2–4]. Rate coding is simple, whereas temporal coding reduces spike counts by lowering event density. Analog inputs can directly drive LIF neurons that fire more for strong signals and sparsely for weak ones [5–7]. Such front ends are used in event-based vision and biomedical systems, where eliminating the ADC improves efficiency [8].

Hardware LIF neurons integrate input current on a capacitor and fire when a threshold is crossed. Their F–I curve is tunable through input-stage transconductance and membrane leak, which control firing rate and spike shape [7]. Effective analog encoders therefore use adjustable gain and leak to support wide dynamic range. Silicon neuron designs focus on low energy per spike, programmable dynamics, and small area [9], often leveraging subthreshold transistor physics to mimic biological membranes for efficient event-driven computation [10, 11].

Many low-power applications require circuits that operate reliably in the subthreshold region [12, 13]. Conventional gate-driven MOS inputs lose range near threshold, whereas bulk-driven transistors maintain channel control and enable rail-to-rail operation in this region [14]. Early work showed that bulk-driven stages can provide gain without forward-biasing the body diode, and later studies demonstrated that combining bulk-driven inputs with subthreshold biasing yields operational transconductance amplifiers (OTA) operating robustly at nanoampere currents [14, 15]. OTAs and transconductors are key blocks in sensor interfaces, enabling functions such as filtering and signal conversion [15]. Bulk-driven transconductors support ultra-low-voltage operation, providing linear V–I conversion at 0.3–0.5 V with rail-to-rail input and tunable transconductance in single-stage or class-AB forms [13–15]. In neuromorphic front ends, these low-voltage linear stages convert input voltages into currents that set an LIF neuron's firing rate, while the leak path defines the integration time constant. This combination enables precise, wide-range analog-to-spike encoding.

This work combines a bulk-driven transconductor front end with an LIF neuron back end operating at 0.5 V. The bulk-driven OTA provides rail-to-rail input range and linear gain in subthreshold, enabling accurate small- and large-signal voltage-to-current encoding. The LIF neuron converts this current into a spike rate proportional to input amplitude, with adjustable gain and leak allowing wide dynamic range and tunable firing behavior. This architecture offers an efficient analog-to-spike encoder suited for energy-constrained neuromorphic sensors, eliminating the need for an ADC and maintaining functionality at very low supply voltage.

## II. Proposed Encoder

Fig. 1 shows the proposed signal-to-spike encoder. The differential input voltage is first converted to a proportional current by a linear transconductor, which drives the LIF neuron. The neuron integrates this current on the membrane capacitor $C_m$, producing higher firing rates for larger input amplitudes and lower rates for smaller ones. Thus, the input amplitude is encoded directly into spike frequency for neuromorphic processing.

### A. Linear Transconductor

In the proposed linear transconductor (highlighted by the green dashed line in Fig. 1), two common-source devices form a tail-less differential pair. Since the circuit targets ultra-low supply operation ($V_{DD}$=0.5 V), all MOS devices are biased in weak inversion. In this region, the drain current depends exponentially on the source-gate voltage, which inherently limits the linearity of conventional voltage-to-current conversion. For a long-channel MOSFET operating in weak inversion and saturation ($V_{SD} \gtrsim 3U_T$), the drain current can be expressed as [14]

$$I_D = I_S \left(\frac{W}{L}\right) \exp\left(\frac{V_{SG} - V_{T0} + (n-1)V_{SB}}{nU_T}\right) \quad (1)$$

where $I_S$ represents $2n\mu C_{ox} U_T^2$, $U_T$ is the thermal voltage with a value of 25 mV at room temperature, $n$ is the slope factor, and $V_{T0}$ is the zero-bias threshold voltage ($V_{SB}$=0). Since a fixed gate bias sets the device operating point, the input signal can be applied through the bulk terminal so that modulation of $V_{SB}$ produces the desired current variation (bulk-driven operation). From (1), the drain current exhibits lower sensitivity to $V_{SB}$ than to $V_{SG}$; for n≈1.2 in a 0.18-μm

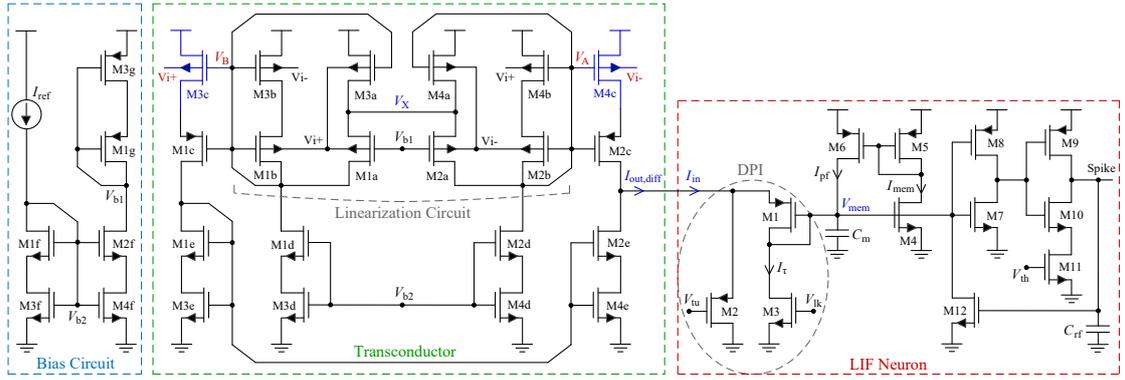

Fig. 1. Proposed signal-to-spike encoder.

CMOS process, this yields approximately $g_{m,bulk} \approx 0.2 \times g_{m,gate}$. Therefore, bulk-driven operation improves linearity at the expense of reduced effective transconductance. According to (1) and the large-signal analysis in [14], the differential output current $I_{out,diff}=(I_{D,4c}-I_{D,3c})$ of the tail-less devices M3c and M4c can be expressed as

$$I_{out,diff} = 2I_{D_{3c}} \times \sinh\left(\overbrace{\frac{V_B - V_A}{2nU_T}}^{\alpha} - \overbrace{\frac{(n-1)(V_{i+} - V_{i-})}{2nU_T}}^{\beta}\right) \quad (2)$$

where $I_{D,3c}=I_{D,4c}$ are the drain currents at the operating point. The voltages $V_B$ and $V_A$ are the differential signals driving the gates of M3c and M4c, respectively, generated by the internal linearization circuit. Based on the Taylor series expansion, $\sinh(x) \approx x$ for $x \ll 1$. Therefore, if the value of $(V_B-V_A)$ approaches $(n-1)(V_{i+}-V_{i-})$, equation (2) becomes highly linearized, although the transconductance $G_m=I_{out,diff}/(V_{i+}-V_{i-})$ is significantly reduced. To achieve this condition, the differential currents between counterpart devices in the linearization circuit are similarly expressed as $I_{D,4a}-I_{D,3a}=2I_{D,3a}\times\sinh(\alpha+\beta)$ and $I_{D,4b}-I_{D,3b}=2I_{D,3b}\times\sinh(\alpha-\beta)$. Next, the summation of these two currents is multiplied by the large-signal resistances seen from nodes A and B, given by $(2nU_T/I_{D,3b}+I_{D,3a})$, which can be expanded as:

$$V_B - V_A = \frac{4nU_T}{I_{D_{3b}} + I_{D_{3a}}} \left( \begin{array}{l} (I_{D_{3b}} + I_{D_{3a}})\cosh(\beta)\sinh(\alpha) + \\ (I_{D_{3b}} - I_{D_{3a}})\cosh(\alpha)\sinh(\beta) \end{array} \right) \quad (3)$$

As observed, the parameter $(I_{D,3b}-I_{D,3a})/(I_{D,3b}+I_{D,3a})$ controls the magnitude of $(V_B-V_A)$, thereby enabling linearization of (2). Linear operation requires the argument of hyperbolic sine function to remain bounded, $|\alpha+\beta|<\varepsilon$, where $\varepsilon$ is dimensionless linearity tolerance parameter, typically chosen within the range $\varepsilon \in [0.03, 0.1]$. By analytical reduction of the large-signal translinear equations and enforcing this boundedness condition, the following sufficient linearity constraint is obtained:

$$\left|\frac{I_{D_{3b}} - I_{D_{3a}}}{I_{D_{3b}} + I_{D_{3a}}}\right| \leq \frac{|\beta| + \varepsilon}{2|\sinh(\beta)|} \quad (4)$$

where $I_{D,i}$ denotes the drain current of the corresponding device at its operating point, and $\beta$ has a linear relationship with the input differential signal, as defined in (2). It follows from (4) that maximum linearity is achieved under symmetric DC biasing of M3a and M3b. Although the proposed circuit effectively cancels the dominant translinear (sinh-based) nonlinearity, residual distortion persists due to second-order device effects such as finite output resistance, mobility degradation, body-effect curvature, channel-length modulation, device mismatch, and parasitic capacitances, which are not captured by the ideal exponential model.

To ensure that the source voltages of all pMOS devices in the linearization circuit track the input signal, the differential input signals are also applied to the bulk terminals of the cascoded devices (M1a, M2a, M1b, and M2b). This approach minimizes body-effect-induced threshold modulation, improves circuit symmetry, and enhances tolerance to process, voltage, temperature (PVT) variations.

B. *LIF Neuron*

The LIF neuron, highlighted by the red dashed line in Fig. 1, is inspired by [7, 11]. The output current of the proposed transconductor is injected into the neuron through a differential pair integrator (DPI) at its input, composed of transistors M1-M3, rather than directly into the membrane node $V_{mem}$. Owing to the cascoded topology, the transconductor exhibits a high output impedance and therefore behaves approximately as a voltage-controlled current source over the intended operating range. As a result, the injected current is only weakly affected by the voltage variation at the neuron input. During the integration phase, the DPI circuit accepts the injected current while the associated node voltage varies within a limited range required to maintain the subthreshold conduction of transistor M1 under a 0.5-V supply. Consequently, the transconductor output remains within its compliance range and the injected current is preserved. Larger voltage excursions occur only during spike generation, when the neuron leaves the integration regime and enters the regenerative spiking phase, where strict linearity is no longer required.

A DPI circuit is used as the front-end integrator, providing a well-defined first-order current-mode integration onto the membrane capacitor $C_m$ with tunable gain and reduced sensitivity to device mismatch [11, 16]. In particular, its gain is controlled by the gate voltage of M2 ($V_{tu}$), while the neuron's leak conductance is adjusted through the bias voltage of M3 ($V_{lk}$). As the input current is integrated onto $C_m$, the membrane voltage $V_{mem}$ gradually increases. When $V_{mem}$ becomes sufficiently large to turn on M4 and generate a noticeable subthreshold current, the positive-feedback loop formed by M4–M6 rapidly increases $V_{mem}$, initiating spike generation. Subsequently, the first inverter (M7-M8) switches its output from $V_{DD}$ to 0, which drives the second inverter (M9–M11) to produce a full-swing output spike and charge the refractory capacitor $C_{rf}$. The additional NMOS device M11 introduces an adjustable threshold $V_{th}$ to the inverter-based comparator, providing flexibility for different applications. After spike generation, M12 turns on and discharges the membrane to ground, resetting the neuron for the next integration cycle set by the new input amplitude.

Since the neuron encodes the amplitude of the input signal into the firing rate of the output spikes, linearity of this conversion is essential to minimize information loss and improve encoding accuracy. Although high linearity was achieved in the proposed transconductor for voltage-to-current conversion, the linearity of the neuron itself is equally critical for accurate current-to-spike transformation. According to (1) and the analysis in [7, 11], applying the translinear principle to the DPI operating in the subthreshold region yields the following approximate membrane dynamics:

$$\tau_m \frac{d}{dt} I_{mem} + I_{mem}\left(1 - \frac{I_{pf}}{I_\tau}\right) = I_{in} \frac{I_{mem}/I_\tau}{1+(I_{mem}/I_g)} \quad (5)$$

where $I_g = nU_T^2 C_{ox}\mu(W_4/L_4)\exp(V_{tu}-V_{T0}/nU_T)$ defines the DPI gain current, $I_{mem}$ represents the membrane state variable, and $\tau_m = nU_T C_m/I_\tau$. Although (5) is a first-order nonlinear differential equation, it can be effectively linearized during integration when the membrane current operates in the window $I_g < I_{mem} < I_\tau$, which is practically achievable through bias design. By choosing $I_g$ sufficiently small, the membrane current exceeds it shortly after reset, driving the DPI into its saturation region and causing the right-hand side of (5) to approach the constant value $(I_{in}I_g/I_\tau)$, assuming the input current varies slowly compared to the neuron firing rate. During the initial phase of integration, when the DPI integrates $I_{in}$ onto $C_m$ and $V_{mem}$ has not yet increased sufficiently to activate M4, $I_{mem}$ remains small and the $I_{pf}$ is negligible. As $V_{mem}$ increases, $I_{mem}$ grows rapidly and enters the region where the DPI effectively limits the charging current, resulting in an approximately constant charging slope of the membrane capacitor. Only when $I_{mem}$ approaches the upper bias-defined boundary does the positive-feedback path dominate, leading to rapid regeneration and spike generation. Therefore, appropriate bias separation between $I_g$, $I_\tau$, and the feedback factor ensures an approximately linear current-to-spike conversion governed by the first-order linear membrane dynamics:

$$\tau_m \frac{d}{dt} I_{mem} + I_{mem} = I_{in} \frac{I_\tau}{I_g} \quad (6)$$

III. EXPERIMENTAL RESULTS

The encoder was designed and fabricated in a TSMC 0.18-μm CMOS process and operates from a 0.5-V supply. Fig. 2 shows the measurement setup. The differential input ($V_{i+}-V_{i-}$) is linearly converted to a single-ended current $I_{in}$ by the transconductor and integrated onto the membrane capacitor $C_m$ by the neuron. Output spikes are buffered to drive the PADs without loading. A differential amplifier in voltage-follower mode monitors the membrane potential, while an inverter chain buffers the spikes. The follower adds an estimated parasitic capacitance of $C_{par}$=70 fF, increasing the effective membrane capacitance to about 0.6 pF. Fig. 3 shows the micrograph of the fabricated encoder, including the neuron, transconductor, and buffers. The neuron and its buffers occupy 71.5μm × 49.7μm, and the transconductor occupies 70.9μm × 54.2μm.

To measure the DC transfer characteristics, the output current of the transconductor was measured using an active current/voltage converter created by an OPA29990 and a 15-MΩ high-precision resistor [15]. As shown in Fig. 4, the reference current was regulated from 2 to 27 nA, as in [13], while the total harmonic distortion (THD) was from 1.76 to 7.2 % for the whole investigated range of the reference currents and for $V_{id}=V_{i+}-V_{i-}$ less than 500 mV. The measured frequency characteristics of the transconductor for 2nA<$I_{ref}$<27nA and $C_L$=20 pF are shown in Fig. 5. The average of the low-frequency gain was around 30 dB, and the unity-gain frequency was in the range of 47–546 Hz, while phase margin was above 85° for all cases. Although measurement with $C_L$<20 pF is not feasible due to pad, probe, and PCB parasitics, the transconductor directly drives the neuron input, which is expected to be on the order of 100-200 fF; therefore, the bandwidth should extend into the tens-of-kHz range.

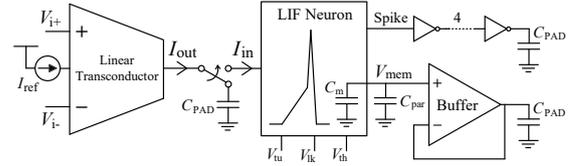

Fig. 2. Circuit configuration used for measurements.

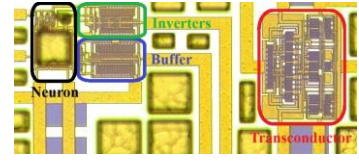

Fig. 3. Microphotograph of the proposed encoder.

To measure the transient behavior of the encoder, control parameters including $I_{ref}$, $V_{tu}$, $V_{lk}$, and $V_{th}$ were set to 1 nA, 0.25 V, 0.1 V, and 0.25 V, respectively. Then, a triangular differential input voltage signal with an amplitude of 0.3 V was applied while simultaneously monitoring the neuron spike output, as shown in Fig. 6. The firing rate follows the input amplitude linearly: it increases with larger inputs and decreases with smaller ones.

To evaluate the DC transfer characteristics of the encoder, $I_{ref}$ was changed to 8 nA and differential input voltage was swept across 0–$V_{DD}$, while the firing rate of the neuron was measured using a digital frequency counter, as shown in Fig. 7. The measured V–F curve exhibits near-linear behavior for $0.1V<V_{in}<0.4V$ with a maximum deviation of less than 5.6 % from the fitted straight line, confirming good linearity over the investigated input range. For larger input amplitudes, the firing rate gradually saturates due to the finite membrane integration and reset dynamics of the neuron, but remains tunable via the control parameters. With appropriate biasing ($V_{lk}$ ↓, $V_{tu}$ ↑, and $V_{th}$ ↑), the upper boundary of the linear firing range exceeds 40 kHz, satisfying $f_{spike}>5\times f_{signal}$ and enabling 8 kHz input bandwidth suitable for biological time-scale.

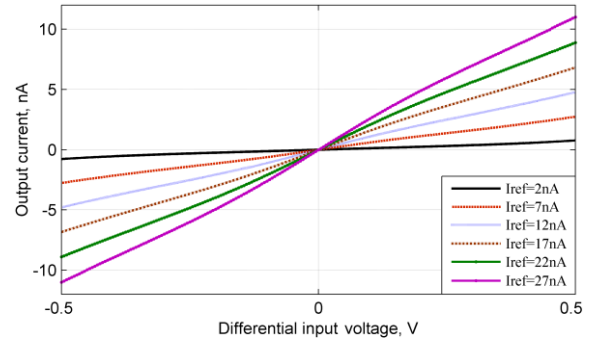

Fig. 4. Measured DC transfer characteristics of the transconductor.

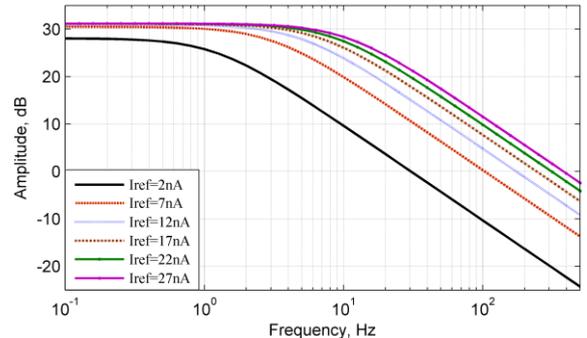

Fig. 5. Measured frequency characteristics of the transconductor.

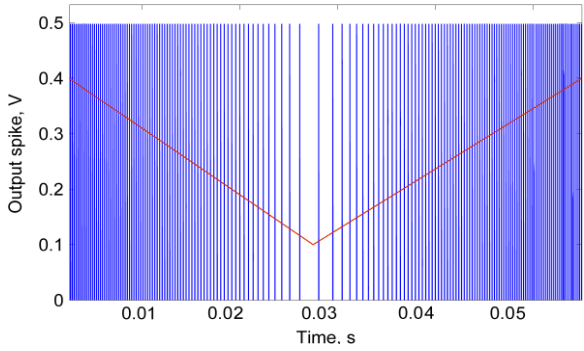

Fig. 6. Measured output spikes for a triangular input voltage signal.

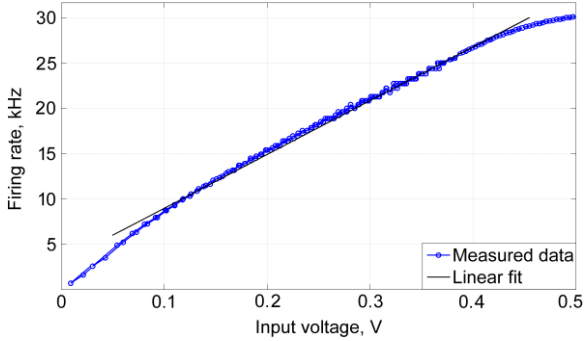

Fig. 7. Measured firing rate versus the amplitude of the input signal.

Table I compares the proposed encoder with other works. This design consumes 22–180 nW power, depending on the neuron's parameters and gain of the transconductor adjusting by $I_{ref}$. It also supports a wide 0.1–0.4 V input voltage range in which the conversion remains linear with less than 5.6 % error while operating from a 0.5-V supply.

## IV. CONCLUSION

This paper presented a low-voltage, low-power linear signal-to-spike encoder suitable for neuromorphic processing. The proposed front-end transconductor converts a differential voltage input into a single-ended current that drives a LIF neuron for spike generation. Implemented in TSMC 0.18 μm CMOS, the circuit achieves linear voltage-to-spike conversion over a 0.1–0.4 V input range while operating from a 0.5-V supply.

## ACKNOWLEDGMENT

This work was partially supported by the European Research Council (ERC) through the European's Union Horizon Europe Research and Innovation Programme under Grant Agreement No 101042585. Views and opinions expressed are however those of the authors only and do not necessarily reflect those of the European Union or the European Research Council. Neither the European Union nor the granting authority can be held responsible for them.

Table I. Specifications of the proposed encoder against other works.

| Parameter | This work | [2] 2023 | [3] 2016 | [4] 2018 |
|---|---|---|---|---|
| Power supply [V] | 0.5 | 1.8 | 0.5 | 0.6 |
| Technology [nm] | 180 | 180 | 180 | 180 |
| Neuron type | LIF | LIF | ADM | BNN |
| Input type | V/Diff | V/Single | V/Diff | V/Diff |
| Linearity error [%] | 5.6 | ≈ 6 | < 6.3 | - |
| I/V gain [nA/V] | 1.56-22 | - | - | - |
| Bandwidth [Hz] | 0-8k | 100-100k | 8-20k | 100-5k |
| Power[1] [nW] | 22-180 | 800 | 430 | 380 |
| Input noise [μV$_{rms}$] | 6.1 | 1.4 | 30 | < 1 |
| Lin. input range, V | 0.1-0.4 | 0.1-0.4 | 0.3-0.4 | ≈ 0.1 |
| Est. area [mm$^2$] | 0.0074 | 0.09 | 0.26 | ≈ 0.07 |

[1]According to [3] and [7].

Meysam Akbari and Erika Covi would like to acknowledge the financial support of the CogniGron research center and the Ubbo Emmius Fund (University of Groningen). The successful completion of this research was supported by the academic resources and research infrastructure provided by the Taiwan Semiconductor Research Institute, National Institutes of Applied Research. We hereby express our sincere gratitude.